\PassOptionsToPackage{hypertexnames=false}{hyperref}
\documentclass[twocolumn,twocolappendix]{aastex701}

\usepackage{amsmath}
\usepackage{placeins}

\makeatletter
\newenvironment{inlinecolumnfigure}
  {\par\addvspace{6pt}\noindent
   \begin{minipage}{\columnwidth}\centering\def\@captype{figure}}
  {\end{minipage}\par\addvspace{6pt}}
\makeatother

\begin{document}
\tighten

\title{Dust Growth in Evolving Filamentary Molecular Clouds: Signatures in Ionization, Resistivity, and Infrared Scattering}
 
\author[orcid=0009-0000-9609-5953]{Hayato Uchimura}
\affiliation{Graduate School of Science and Engineering, Kagoshima University, 1-21-35 Korimoto, Kagoshima 890-0065, Japan}
\email{k4306527@kadai.jp}

\author[orcid=0000-0002-4093-6925]{Yoshiaki Misugi}
\affiliation{Faculty of Science and Engineering, Kyushu Sangyo University, 2-3-1 Matsukadai, Fukuoka 813-8503, Japan}
\email{misugi@ip.kyusan-u.ac.jp}

\author[orcid=0000-0001-6273-805X]{Daisuke Takaishi}
\affiliation{Graduate School of Science and Engineering, Kagoshima University, 1-21-35 Korimoto, Kagoshima 890-0065, Japan}
\email{takaishi@ibe.kagoshima-u.ac.jp}

\author[orcid=0009-0009-1967-7592]{Taishi Tsuchiyama}
\affiliation{Graduate School of Science and Engineering, Kagoshima University, 1-21-35 Korimoto, Kagoshima 890-0065, Japan}
\email{k4319324@kadai.jp}

\author[orcid=0009-0006-2626-1454]{Tomoki Tokunaga}
\affiliation{Graduate School of Science and Engineering, Kagoshima University, 1-21-35 Korimoto, Kagoshima 890-0065, Japan}
\email{k9493595@kadai.jp}

\author[orcid=0009-0006-3692-213X]{Rintaro Shibakawa}
\affiliation{Graduate School of Science and Engineering, Kagoshima University, 1-21-35 Korimoto, Kagoshima 890-0065, Japan}
\email{k1629504@kadai.jp}

\author[orcid=0000-0001-6738-676X]{Yusuke Tsukamoto}
\affiliation{Department of Physics, Faculty of Science and Engineering, Konan University, 8-9-1 Okamoto, Higashinada-ku, Kobe 658-8501, Japan}
\email{tsukamoto@konan-u.ac.jp}

\begin{abstract}
Studies of dust growth in molecular clouds often prescribe the collapse history, leaving unclear how magnetic regulation of core formation shapes dust evolution and its observable and non-ideal MHD signatures. We address this problem by performing one-zone dust evolution and ionization calculations along time-dependent density and magnetic-field histories extracted from three-dimensional non-ideal MHD simulations of core formation through filament fragmentation. A stronger initial magnetic field delays contraction and thereby gives dust grains more time to grow at a given density. In our models, this delayed contraction leads to mass-weighted mean dust sizes of $1.2\,\mu\mathrm{m}$ and $3.1\,\mu\mathrm{m}$ at $n_{\mathrm{H}}=10^6\,\mathrm{cm^{-3}}$ for $B_{\mathrm{ini}}=20$ and $50\,\mu\mathrm{G}$, respectively. The associated depletion of very small dust grains reduces the adsorption of charged particles onto dust grain surfaces and lowers the conductivity: relative to models without dust growth, the ionization fraction increases from $\sim10^{-9}$ to $\sim10^{-8}$, the ambipolar resistivity increases, and the ion--neutral drift velocity rises to several $\mathrm{m\,s^{-1}}$ in the weak-magnetic-field model and approximately $10\,\mathrm{m\,s^{-1}}$ in the strong-magnetic-field model. These drift velocities remain below the observationally suggested range of $30$--$100\,\mathrm{m\,s^{-1}}$, indicating that still stronger magnetic fields may be required. The evolved dust populations also attain single-scattering albedos of order $\omega_\lambda\sim0.8$ at $3.6$--$4.5\,\mu\mathrm{m}$, with the stronger field shifting the onset of efficient infrared scattering toward lower densities. These results demonstrate that the magnetically controlled collapse timescale, rather than density alone, links dust growth to ionization, ambipolar diffusion, and infrared scattering in prestellar cores.
\end{abstract}


\small

\section{Introduction}

Whether dust grains grow appreciably during the molecular cloud and prestellar-core phases remains an important open question. Directly determining the dust size distribution inside cold, dense molecular gas is difficult, and the observational evidence for dust growth in these environments is therefore largely indirect.

One of the most widely discussed indications of dust growth is ``coreshine'', mid-infrared light scattered by dust grains in dense molecular cloud cores. The ``coreshine'' phenomenon has been detected toward many dense cores at wavelengths of $3.6$--$4.5\,\mu\mathrm{m}$ \citep{2010A&A...511A...9S,2010Sci...329.1622P,2014A&A...564A..96S,2015A&A...582A..70S}. Radiative-transfer models commonly require dust grains with sizes of order $1\,\mu\mathrm{m}$ to reproduce efficient scattering at these wavelengths \citep{2013A&A...559A..60A}. Such dust grains are substantially larger than the upper cutoff of the standard MRN size distribution for diffuse interstellar dust grains \citep{1977ApJ...217..425M}, suggesting that dust growth may already occur before protostar formation.

The ``coreshine'' phenomenon alone, however, does not uniquely demonstrate that dust grains have grown in-situ during dense-core formation. Its surface brightness depends not only on the dust size distribution, but also on the incident radiation field, dust composition, scattering phase function, optical depth, density structure, and line-of-sight geometry. Moreover, the inferred large dust grains may reflect either an initially present component of large dust grains or the earlier dynamical history of the cloud \citep{2013A&A...559A..60A,2014A&A...564A..96S}. The ``coreshine'' phenomenon therefore indicates the presence of dust grains capable of efficient mid-infrared scattering, but does not by itself establish when or where those dust grains grew.

A complementary way to investigate dust growth is through its influence on the coupling between neutral gas and magnetic fields. Very small dust grains dominate the total dust surface area and can make an important contribution to the electrical conductivity. Their depletion by coagulation reduces the adsorption of charged particles onto dust surfaces and their subsequent recombination, and removes charged dust grains that contribute to the conductivity perpendicular to the magnetic field. Dust growth can consequently change both the gas-phase ion abundance and the Hall and ambipolar resistivities, thereby altering the gas--magnetic-field coupling and the characteristic ion--neutral drift velocity \citep{1979ApJ...232..729E,1986MNRAS.218..663N,1990MNRAS.243..103U,2016MNRAS.460.2050Z,2021A&A...649A..50M,2022ApJ...934...88T}.

Using one-zone calculations under fixed density and magnetic-field conditions, \citet{2026ApJ...999...79F} quantified this connection. At dense-core densities, they found that a characteristic ion--neutral drift velocity of order $100\,\mathrm{m\,s^{-1}}$ required both substantial dust growth and a magnetic-field strength of order $200\,\mu\mathrm{G}$. If the initial MRN-like dust grain population was retained, a field strength of $\gtrsim1\,\mathrm{mG}$ was instead required to produce a comparable drift velocity. Thus, for magnetic fields typical of dense cores, suppressing dust growth leaves the predicted ion--neutral drift velocity much smaller than in models where very small dust grains are depleted.

This result identifies ion--neutral drift, which at first sight appears unrelated to dust evolution, as a potential indirect diagnostic of dust growth. An observed ion--neutral drift velocity difference can depend not only on the gas density and magnetic-field strength, but also on the underlying dust size distribution through the magnetic resistivities. Dust evolution must therefore be considered when ion--neutral drift is used to diagnose magnetic coupling in star-forming gas.

Recent comparisons of ionic and neutral molecular lines have begun to constrain ion--neutral drift velocities in both prestellar cores and protostellar envelopes. Toward the L1544 prestellar core, \citet{2026arXiv260522541A} reported a mean velocity difference of approximately $0.05\,\mathrm{km\,s^{-1}}$ between N$_2$D$^+$ and para-NH$_2$D and interpreted it as a signature of ambipolar diffusion. On a scale of approximately $100\,\mathrm{au}$ in the Class~0 protostar B335, \citet{2018A&A...615A..58Y} obtained an upper limit of approximately $0.3\,\mathrm{km\,s^{-1}}$ on the difference between ion and neutral infall velocities. These measurements and constraints are affected by tracer selection, line-of-sight averaging, projection, and radiative transfer, but they reinforce the need to include dust evolution when interpreting ion--neutral velocity differences.

The fixed conditions adopted by \citet{2026ApJ...999...79F}, however, do not describe dust growth in a dynamically evolving 
cloud. During core formation, both the density and magnetic-field strength change continuously, while coagulation and ice accretion require a finite residence time at elevated densities to modify the dust size distribution. The timescale for substantial dust growth can be comparable to, or longer than, the dynamical evolution time of dense gas \citep{1993A&A...280..617O,2009A&A...502..845O,2009MNRAS.399.1795H,2026ApJ...999...79F}. It is therefore unclear whether dust grains can grow far enough along a realistic collapse history to produce the scattering properties associated with ``coreshine'' and to modify the magnetic resistivities and ion--neutral drift appreciably.

The filamentary structure of molecular clouds provides an important context for this timescale problem. Observations with the \textit{Herschel} Space Observatory revealed ubiquitous filamentary structures in nearby molecular clouds and showed that dense cores are preferentially associated with dense, gravitationally unstable filaments \citep{Arzoumanian_2011,Arzoumanian2019,Andre2014,Pineda2023}. The relevant dynamical background for prestellar dust growth may therefore be core formation through filament contraction and fragmentation rather than an isolated spherical free-fall collapse.

Filamentary evolution may provide more time for dust growth before dense-core conditions are reached. Analytic models of finite self-gravitating filaments show that their global collapse time can exceed the spherical free-fall time at the same density by an aspect-ratio-dependent factor \citep{2012ApJ...756..145P}. In addition, a filament can remain approximately supported in the radial direction while unstable perturbations grow along its axis, so core formation through fragmentation need not follow a prompt spherical free-fall trajectory. A magnetic field perpendicular to the filament can further reduce the growth rate of fragmentation and suppress contraction across the field, while magnetic fields and turbulent motions generally regulate the density evolution and the time available for dust growth \citep{2017ApJ...848....2H,2024ApJ...963..106M}. These considerations motivate the hypothesis that the filamentary mode of core formation provides conditions favorable for dust growth and thereby amplifies its signatures in ion--neutral drift and infrared scattering.

In this study, we test this hypothesis by replacing the fixed density and magnetic-field conditions adopted by \citet{2026ApJ...999...79F} with representative time-dependent histories extracted from three-dimensional non-ideal MHD simulations of core formation through the fragmentation of magnetized, turbulent molecular filaments. Along these histories, we perform post-processing one-zone calculations of dust grain coagulation and ice accretion. We then evaluate the resulting ionization balance, ambipolar resistivity, characteristic ion--neutral drift velocity, and mid-infrared scattering properties, including the single-scattering albedo calculated using OpTool \citep{2021ascl.soft04010D}. By comparing histories with different initial magnetic-field strengths, we examine whether magnetically regulated filament evolution provides sufficient time for appreciable dust growth and how the resulting dust evolution may appear in ion--neutral drift and infrared scattering diagnostics. The present calculation is one-way post-processing: the resistivity changes caused by dust growth are not fed back into the three-dimensional gas and magnetic-field evolution, and we do not perform radiative-transfer calculations of the ``coreshine'' surface brightness.

This paper is organized as follows. In Section~\ref{sec:2}, we describe the ionization and resistivity calculations, the background gas evolution, the dust evolution model, and the optical-property calculation for dust grains. In Section~\ref{sec:3}, we present the dust growth along the filament-evolution histories and its effects on the ionization balance, ambipolar resistivity, characteristic ion--neutral drift velocity, and infrared scattering properties. In Section~\ref{sec:4}, we discuss the implications and limitations of our results.

\section{Methods}
\label{sec:2}

\subsection{Ionization Equilibrium, Magnetic Resistivity, and Ion--neutral Drift}
\label{subsec:calc_eta}

We calculate the ion and electron abundances and the charge distribution of dust grains assuming equilibrium among cosmic-ray ionization, gas-phase recombination, adsorption of charged particles onto dust grains, and dust grain charging. We follow the analytical method of \citet{2022ApJ...934...88T}, as adopted for cold molecular cloud conditions by \citet{2026ApJ...999...79F}; see these studies for the detailed equations and rate coefficients. 
At $T=10\,{\rm K}$, the initial MRN distribution
($a_{\max}=0.25\,\mu{\rm m}$) has $\tau_{\max}\simeq0.15$,
where $\tau\equiv a k_{\rm B}T/e^2$, so the
$Z=-1,0,+1$ approximation is well justified initially.
However, $\tau=1$ corresponds to $a\simeq1.67\,\mu{\rm m}$,
and the largest grains produced in our calculations reach
$\tau\simeq16.5$, formally outside this regime.
We therefore tested the general all-$\tau$ approximation of
\citet{2021A&A...649A..50M}.
The changes in the electron and ion abundances remain below $1\%$,
while grains with $\tau>1$ contribute at most $0.03\%$ of the net grain charge.
We thus retain the three-charge-state approximation in our fiducial
calculations.
From the resulting abundances of ions, electrons, and charged dust grains, we calculate the ambipolar resistivity $\eta_{\mathrm{A}}$, following the same formulation.

We estimate a characteristic ion--neutral drift velocity from the ambipolar resistivity as
\begin{equation}
v_{\mathrm{drift}}\sim\frac{\eta_{\mathrm{A}}}{L_B}
\simeq\frac{\eta_{\mathrm{A}}}{\lambda_{\mathrm{J}}},
\qquad
\lambda_{\mathrm{J}}=c_{\mathrm{s}}\left(\frac{\pi}{G\rho_{\mathrm{g}}}\right)^{1/2},
\label{eq:vdrift_estimate}
\end{equation}
where $L_B$ is the characteristic scale of magnetic-field variation, which we approximate by the Jeans length $\lambda_{\mathrm{J}}$ \citep{2020A&A...641A..39S,2024A&A...690A..23V,2026ApJ...999...79F}. Thus, $v_{\mathrm{drift}}$ in this work is a characteristic estimate rather than a drift velocity measured directly from the three-dimensional simulation.

\subsection{Gas Evolution}
\label{subsec:gas_evolution}
The gas-density and magnetic-field histories used in the dust evolution
calculations are extracted from three-dimensional, self-gravitating,
isothermal non-ideal MHD simulations of a fragmenting filament
(Misugi et al., in preparation). The simulations include ambipolar
diffusion and employ the Godunov smoothed particle
magnetohydrodynamics method \citep{Iwasaki2011,Iwasaki2013}. 
The initial filament is represented by approximately
$1.5\times10^{7}$ SPH particles, with a particle mass of
$3.125\times10^{-6}\,M_{\odot}$. The
Lagrangian nature of SPMHD allows us to directly trace the density and
magnetic-field histories of the gas incorporated into a forming core.

The initial filament has the hydrostatic-equilibrium density profile \citep{1964ApJ...140.1056O}
\begin{equation}
\begin{aligned}
\rho_{\mathrm{g}}(r)
&=\rho_{\mathrm{c}0}\left[1+\left(\frac{r}{H_0}\right)^2\right]^{-2},\\
H_0
&=\left(\frac{2c_{\mathrm{s}}^2}{\pi G\rho_{\mathrm{c}0}}\right)^{1/2}
=0.05~\mathrm{pc},
\end{aligned}
\label{eq:inidenpro}
\end{equation}
where $r$ is the cylindrical radius and $\rho_{\mathrm{c}0}$ is the initial
central density. We adopt $c_{\mathrm{s}}=0.2~\mathrm{km\,s^{-1}}$, corresponding
to $T_{\mathrm{g}}=10~\mathrm{K}$; the resulting critical line mass is
$18~M_\odot~\mathrm{pc^{-1}}$. The simulated filament segment is
$1.6~\mathrm{pc}$ long. A turbulent velocity field with a Kolmogorov
spectrum and velocity dispersion $\sigma_{\mathrm{turb}}=2c_{\mathrm{s}}$ is imposed
initially. We consider $B_{\mathrm{ini}}=20$ and $50~\mu\mathrm{G}$ at
$n_{\mathrm{H}}\simeq5\times10^4~\mathrm{cm^{-3}}$, with the magnetic field
oriented perpendicular to the filament axis. The corresponding
dimensionless mass-to-flux ratios are $\mu=3.1$ and $1.3$
\citep{Tomisaka1988}; hereafter, these models are referred to as the
weak- and strong-magnetic-field models, respectively.
\begin{figure*}
    \centering
    \hspace*{-0.3cm}
    \includegraphics[scale=0.75]{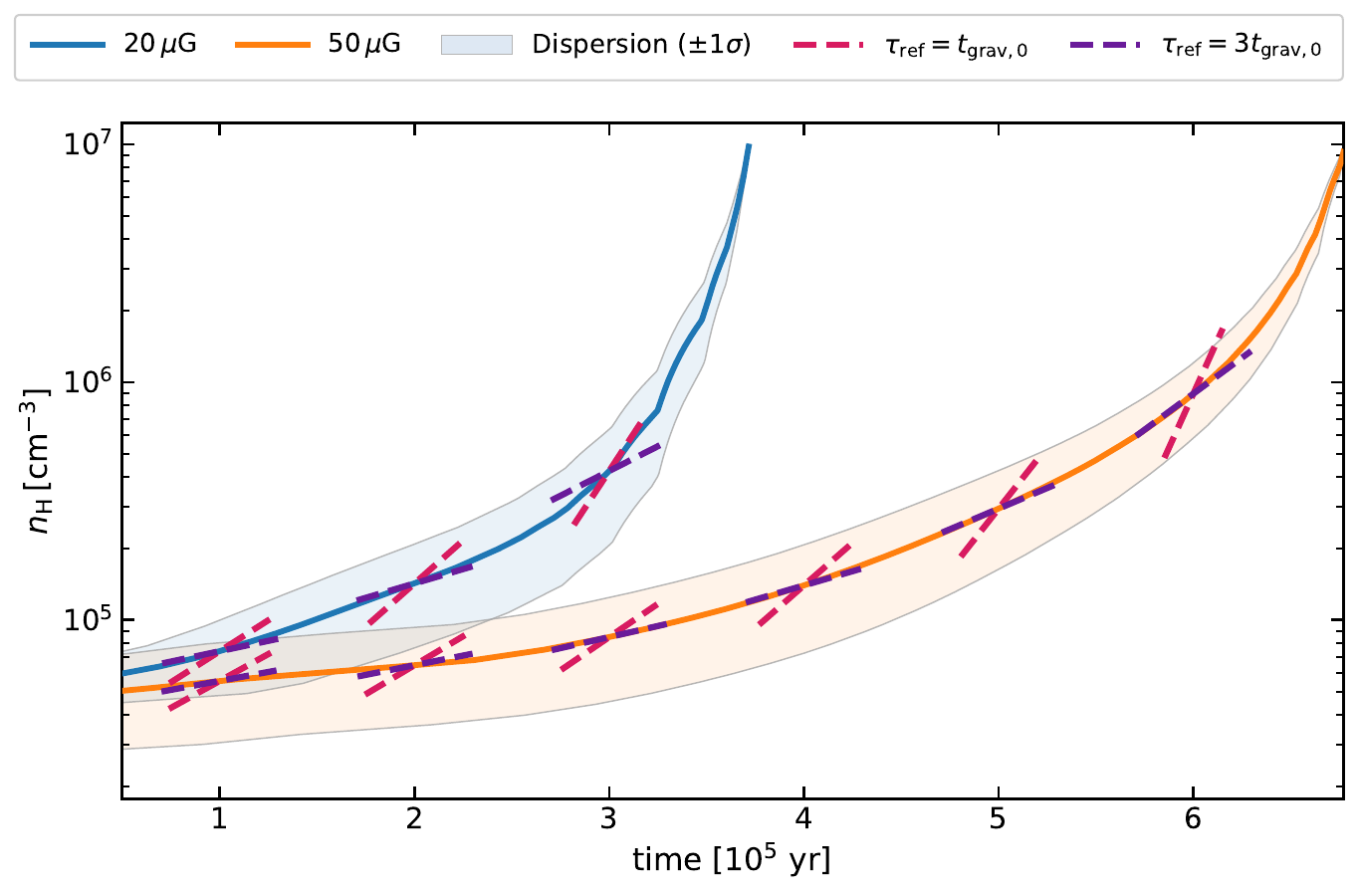}
    \caption{Time evolution of the number density of hydrogen nuclei $n_{\mathrm{H}}$ for the $\mathcal{M}=2.0$ models. The blue and orange solid lines show the mean values for initial magnetic-field strengths of $20\,\mu\mathrm{G}$ and $50\,\mu\mathrm{G}$, respectively, and the shaded regions indicate the corresponding $\pm1\sigma$ dispersion ranges. The magenta and purple dashed lines show reference density-growth guides obtained by assuming $d n_{\mathrm{H}}/dt=n_{\mathrm{H}}/t_{\mathrm{grav},0}$ and $d n_{\mathrm{H}}/dt=n_{\mathrm{H}}/(3t_{\mathrm{grav},0})$, respectively, where $t_{\mathrm{grav},0}=(4\pi G\rho_{\mathrm{g},0})^{-1/2}$ is evaluated from the mean density at each reference time $t_0$, with $\rho_{\mathrm{g},0}=\mu_{\mathrm{H}}m_{\mathrm{p}}\overline{n}_{\mathrm{H}}(t_0)$ and $\mu_{\mathrm{H}}=1.4$. The value of $t_{\mathrm{grav},0}$ is held fixed along each reference guide.}
    \label{fig:nH-time}
\end{figure*}
At the final snapshot, where the maximum gas density reaches
$\rho_{\rm g,max}\sim10^{-14}\,{\rm g\,cm^{-3}}$,
we select the high-density SPH particles using only the density
criterion, $n_{\rm H}\simeq10^{7}\,{\rm cm^{-3}}$, and trace the
same particles backward to the initial state.
The selected region subsequently continues to collapse and develops a single density peak. In contrast, the gas between neighboring cores has a much lower density of \(\sim10^4\,{\rm cm^{-3}}\). Because of this large density contrast, the selected high-density gas is clearly separated from the lower-density inter-core material. Thus, our selection does not depend sensitively on the precise choice of clump-finding or core-definition method. For each model, their ensemble-averaged density and
magnetic-field histories, together with the corresponding mean
$\pm1\sigma$ histories, are used as the prescribed backgrounds for the
dust evolution calculations.

\subsection{Dust Evolution and Post-processing}
\label{subsec:dust_evolution}

We calculate dust evolution along the representative gas-density and
magnetic-field histories extracted from the SPMHD simulations described in
Section~\ref{subsec:gas_evolution}. Our treatment follows
\citet{2026ApJ...999...79F}, based on the dust size distribution formalism of
\citet{2019MNRAS.482.2555H}. The dust mass distribution
$\rho_{\mathrm{d}}(m,t)$ evolves as
\begin{equation}
\frac{\partial \rho_{\mathrm{d}}}{\partial t}
=
\left[\frac{\partial \rho_{\mathrm{d}}}{\partial t}\right]_{\mathrm{coag}}
+
\left[\frac{\partial \rho_{\mathrm{d}}}{\partial t}\right]_{\mathrm{acc}}
+
\rho_{\mathrm{d}}\frac{{\mathrm{d}}\ln\rho_{\mathrm{g}}}{{\mathrm{d}}t}.
\label{eq:dust_evolution_total}
\end{equation}
The coagulation and accretion operators, including depletion of the gas-phase
key species, are identical to those of \citet{2026ApJ...999...79F}; see that
work for their equations and numerical implementation. The last term, which
is absent from their fixed-density calculations, accounts for compression
along the time-dependent gas history. We consider the accretion of oxygen as water ice onto grain surfaces, so that the growing grains are expected to develop ice mantles.
Relative velocities between dust grains arise from Brownian motion and
turbulence \citep{2007A&A...466..413O}.
As in \citet{2026ApJ...999...79F}, all collisions between dust grains are assumed
to lead to coagulation, while bouncing and fragmentation are neglected.

\begin{figure}
  \centering
  \hspace{-0.9cm}
  \includegraphics[scale=0.45]{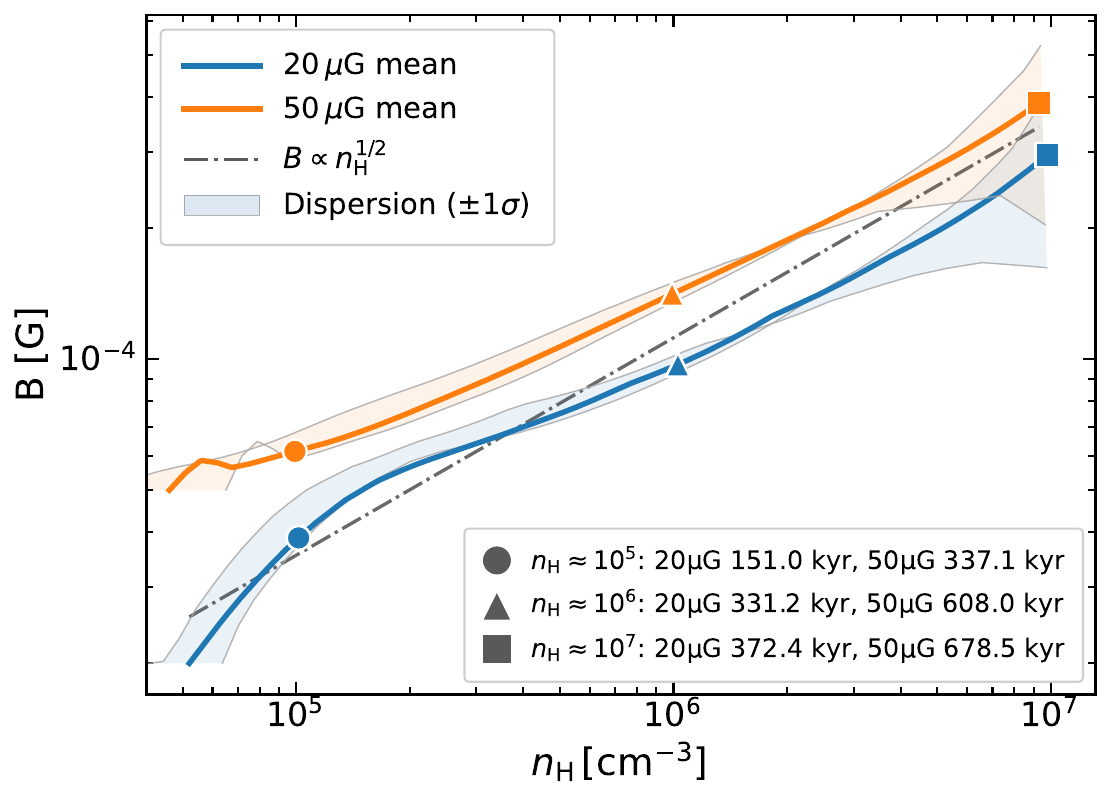}
  \caption{Magnetic field strength $B$ as a function of the number density of hydrogen nuclei $n_{\mathrm{H}}$ for the models with initial magnetic-field strengths of $20\,\mu\mathrm{G}$ and $50\,\mu\mathrm{G}$. The blue and orange solid lines show the mean evolutionary tracks, and the shaded bands indicate the dispersion range ($\pm1\sigma$). The gray dot-dashed line denotes the reference scaling $B \propto n_{\mathrm{H}}^{1/2}$. The circle, triangle, and square symbols mark the epochs at which the mean density reaches $n_{\mathrm{H}}\approx10^{5}$, $10^{6}$, and $10^{7}\,\mathrm{cm^{-3}}$, respectively, and the corresponding times are listed in the legend.}
  \label{fig1}
\end{figure}

The calculation starts from an MRN size distribution,
$n(a)\propto a^{-3.5}$ for $0.005\leq a/(\mu\mathrm{m})\leq0.25$
\citep{1977ApJ...217..425M}, represented by 256 logarithmic bins spanning
$0.003$--$30\,\mu\mathrm{m}$. We adopt $T_{\mathrm{g}}=10\,\mathrm{K}$, a turbulent
Mach number $\mathcal{M}=2.0$, and otherwise use the fiducial dust growth
parameters of \citet{2026ApJ...999...79F}, including a material density for icy dust grains
of $\rho_{\mathrm{mat}}=1.0\,\mathrm{g\,cm^{-3}}$. For the ionization and
resistivity calculations, we set
$\zeta_{\mathrm{CR}}=1.0\times10^{-17}\,\mathrm{s^{-1}}$ and take $\mathrm{H_3^+}$
as the representative gas-phase ion
\citep{2002ApJ...565..344C,2012ApJ...753...29T}.

For comparison, we also calculate models without dust growth in which the dust size
distribution is held at its initial MRN form. In the background SPMHD
simulations, the ambipolar resistivity is evaluated using this unevolved MRN
distribution and the same cosmic-ray ionization rate. Dust evolution is
included only through one-way post-processing: the resulting resistivity
changes are not fed back into the SPMHD evolution. For the infrared-scattering analysis, we repeat the dust evolution calculation using the material densities of the DSHARP mixture \citep{2018ApJ...869L..45B} and astronomical silicate \citep{2003ApJ...598.1017D} and calculate the absorption and scattering opacities of compact spherical dust grains with OpTool \citep{2021ascl.soft04010D}.

\begin{figure*}[t]
  \centering
  \hspace*{-0.3cm}
  \includegraphics[scale=0.6]{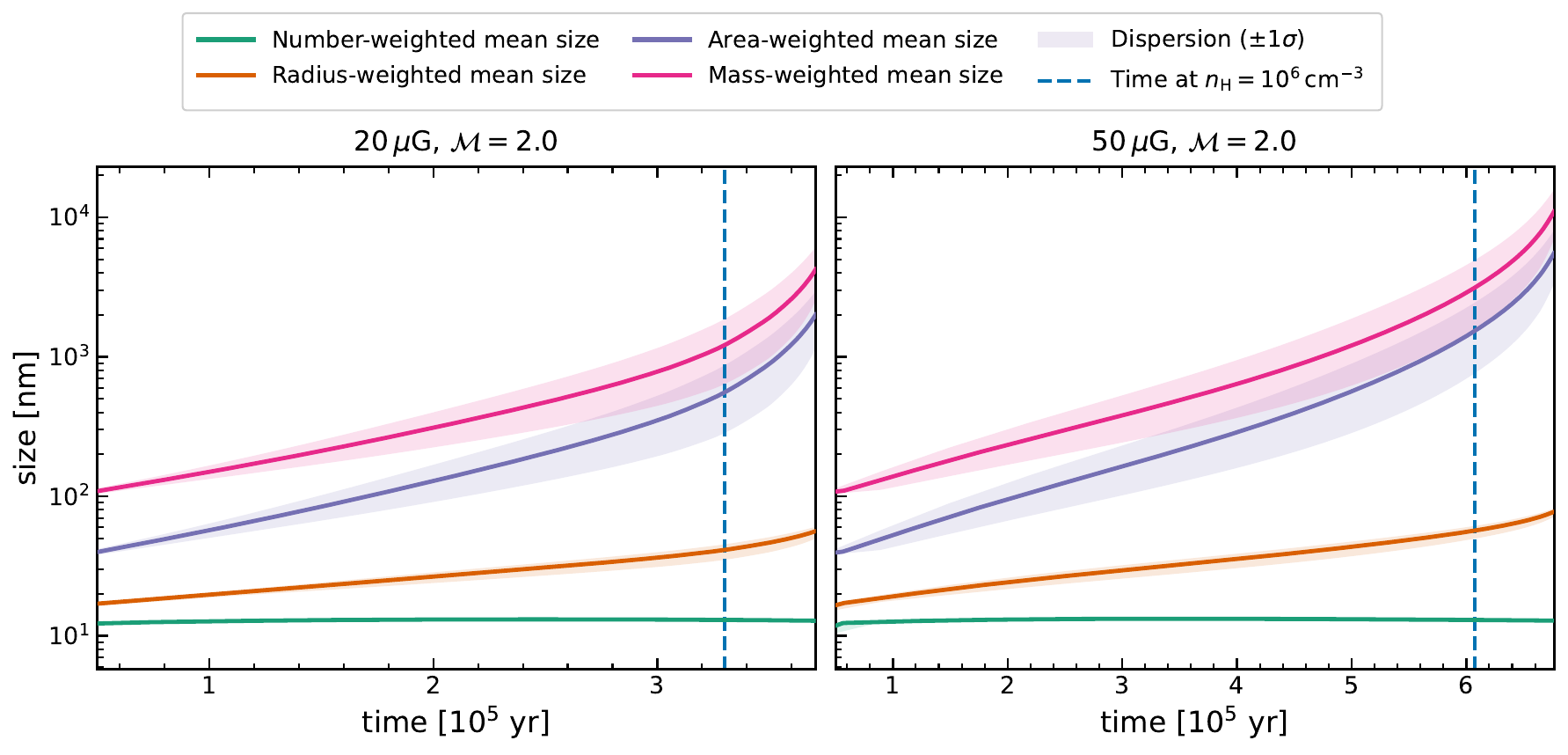}
  \caption{Time evolution of weighted mean dust sizes in the weak- and strong-magnetic-field models (left and right, respectively). The green, orange, blue, and magenta solid lines show the number-, radius-, area-, and mass-weighted mean dust sizes, respectively. The shaded bands indicate the dispersion ranges ($\pm1\sigma$) around the mean evolution. The blue dashed vertical lines mark, for each model, the time at which the mean number density of hydrogen nuclei reaches $n_{\rm H}=10^6\,\mathrm{cm}^{-3}$.}
  \label{fig2}
\end{figure*}

\section{Results}
\label{sec:3}

\subsection{Collapse Timescale}
\label{sec:density-evolution}

Figure~\ref{fig:nH-time} shows the time evolution of the number density of hydrogen nuclei $n_{\mathrm{H}}$ for the weak- and strong-magnetic-field models. The reference density-growth guides are obtained by assuming $d n_{\mathrm{H}}/dt=n_{\mathrm{H}}/t_{\mathrm{grav},0}$ and $d n_{\mathrm{H}}/dt=n_{\mathrm{H}}/(3t_{\mathrm{grav},0})$, respectively. 
Here, $t_{\mathrm{grav},0}$ is defined as
$t_{\mathrm{grav},0}=(4\pi G\rho_{\rm g,0})^{-1/2}$,
where $\rho_{\rm g,0}$ is the mean gas mass density at each reference time $t_0$.
This timescale corresponds to the inverse of the characteristic
gravitational growth-rate scale.
The value of $t_{\mathrm{grav},0}$ is held fixed along each reference guide.
These guides provide benchmarks for comparing the characteristic timescale
of the simulated density evolution with the corresponding reference
gravitational timescale.

In both models, the mean density increases on a timescale longer than
the corresponding reference gravitational timescale
$t_{\mathrm{grav},0}$,indicating that the density evolution is slower than the
reference growth represented by $t_{\mathrm{grav},0}$.
The contraction is slower in the strong-magnetic-field model than in the
weak-magnetic-field model.
The density-growth timescale in the former is approximately
$3t_{\mathrm{grav},0}$ over a wide density range.
This behavior is qualitatively consistent with the linear
stability analysis of \citet{2017ApJ...848....2H}, who showed that a
magnetic field perpendicular to a filamentary cloud reduces the growth rate
of fragmentation and suppresses compression perpendicular to the magnetic
field.
The slower contraction allows the gas to remain longer in each density
regime, providing more time for coagulation and surface accretion.
This longer residence time therefore plays an important role
in the dust growth discussed in Section~\ref{sec:dust-growth}.

  \begin{figure*}[!t]
  \centering
  \hspace*{-0.3cm}
  \includegraphics[scale=0.6]{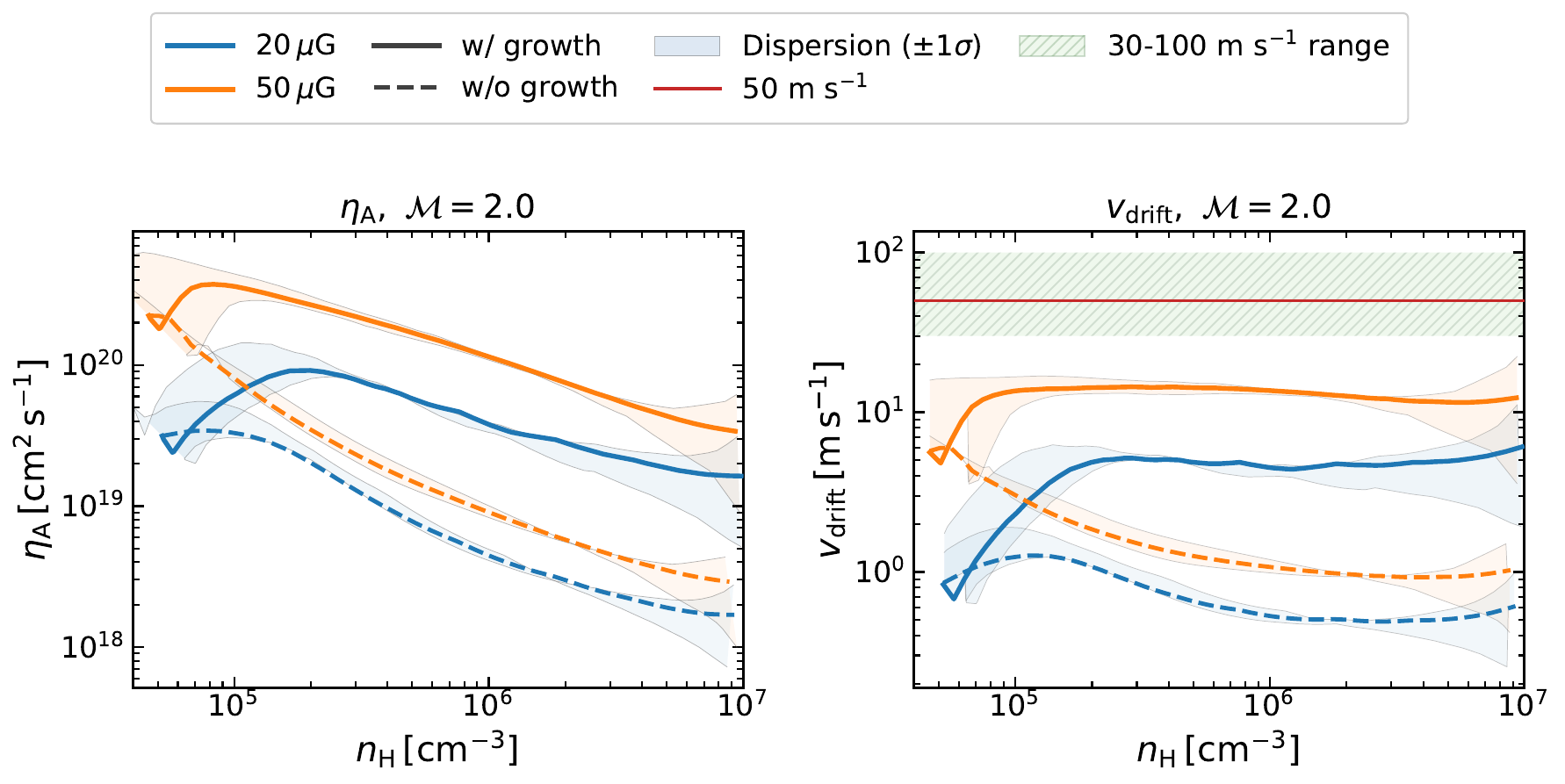}
  \caption{Ambipolar resistivity $\eta_{\mathrm{A}}$ (left) and characteristic ion--neutral drift velocity $v_{\mathrm{drift}}$ (right) as functions of the number density of hydrogen nuclei $n_{\mathrm{H}}$. Blue and orange lines correspond to the weak- and strong-magnetic-field models, respectively. Solid and dashed lines show the models with dust growth and the models without dust growth, respectively, and the shaded bands indicate the dispersion ranges ($\pm1\sigma$). In the right panel, the red horizontal line marks $50\,\mathrm{m\,s^{-1}}$, and the green hatched region indicates the observationally suggested range of $30$--$100\,\mathrm{m\,s^{-1}}$.}
  \label{fig:eta-drift}
\end{figure*}

\subsection{Gas Evolution}

Figure~\ref{fig1} shows the relation between the magnetic field strength $B$
and the number density of hydrogen nuclei $n_{\mathrm{H}}$ for the
weak- and strong-magnetic-field models.

Overall, the evolutionary tracks of both models are approximately consistent with $B\propto n_{\mathrm{H}}^{1/2}$. Modest deviations from this scaling occur during the evolution: the relation is somewhat shallower at $n_{\mathrm{H}}\sim10^{5}$--$10^{6}~\mathrm{cm^{-3}}$, and the strong-magnetic-field model shows a small temporary decrease in $B$ at low densities.

\subsection{Dust Growth}
\label{sec:dust-growth}

Figure~\ref{fig2} shows the time evolution of the weighted mean dust sizes in the weak- and strong-magnetic-field models.

The inferred dust growth depends strongly on the adopted weighting: the number-weighted mean size remains at approximately $10~\mathrm{nm}$ in both models, indicating that small dust grains continue to dominate in number. In contrast, the area-weighted and mass-weighted mean sizes increase substantially, showing that the dust surface area and dust mass shift toward larger dust grains. At $n_{\mathrm{H}}=10^{6}~\mathrm{cm^{-3}}$, the area-weighted mean sizes are $\sim0.56$ and $\sim1.5~\mu\mathrm{m}$, and the mass-weighted mean sizes are $\sim1.2$ and $\sim3.1~\mu\mathrm{m}$, in the weak- and strong-magnetic-field models, respectively. The larger dust grains in the strong-magnetic-field model at the same density can be attributed to its slower contraction, which provides more time for dust growth in each density regime.

\subsection{Ambipolar Resistivity and Ion--neutral Drift Velocity}
\label{sec:resis}

Figure~\ref{fig:eta-drift} shows the ambipolar resistivity $\eta_{\mathrm{A}}$ and the ion--neutral drift velocity $v_{\mathrm{drift}}$ as functions of $n_{\mathrm{H}}$ for the weak- and strong-magnetic-field models with dust growth and the corresponding models without dust growth.

In both models, $\eta_{\mathrm{A}}$ generally decreases with increasing $n_{\mathrm{H}}$, but remains systematically larger with dust growth than without it because the depletion of very small dust grains lowers the conductivity. Since $v_{\mathrm{drift}}\sim\eta_{\mathrm{A}}/\lambda_{\mathrm{J}}$, this enhancement of $\eta_{\mathrm{A}}$ also produces larger ion--neutral drift velocities. At high densities, both $\eta_{\mathrm{A}}$ and the Jeans length $\lambda_{\mathrm{J}}$ scale approximately as $n_{\mathrm{H}}^{-1/2}$, making $v_{\mathrm{drift}}$ only weakly dependent on density. With dust growth, $v_{\mathrm{drift}}$ reaches several $\mathrm{m\,s^{-1}}$ in the weak-magnetic-field model and approximately $10\,\mathrm{m\,s^{-1}}$ in the strong-magnetic-field model, but remains below the observationally suggested range of $30$--$100\,\mathrm{m\,s^{-1}}$ toward L1544 \citep{2026arXiv260522541A}. Reproducing the observed drift velocity may therefore require stronger magnetic fields at the dense-core stage than those realized in the present $B_{\mathrm{ini}}=20$--$50\,\mu\mathrm{G}$ models.

\subsection{Ionization Fraction}

\begin{figure}[!t]
  \centering
  \includegraphics[width=0.95\columnwidth]{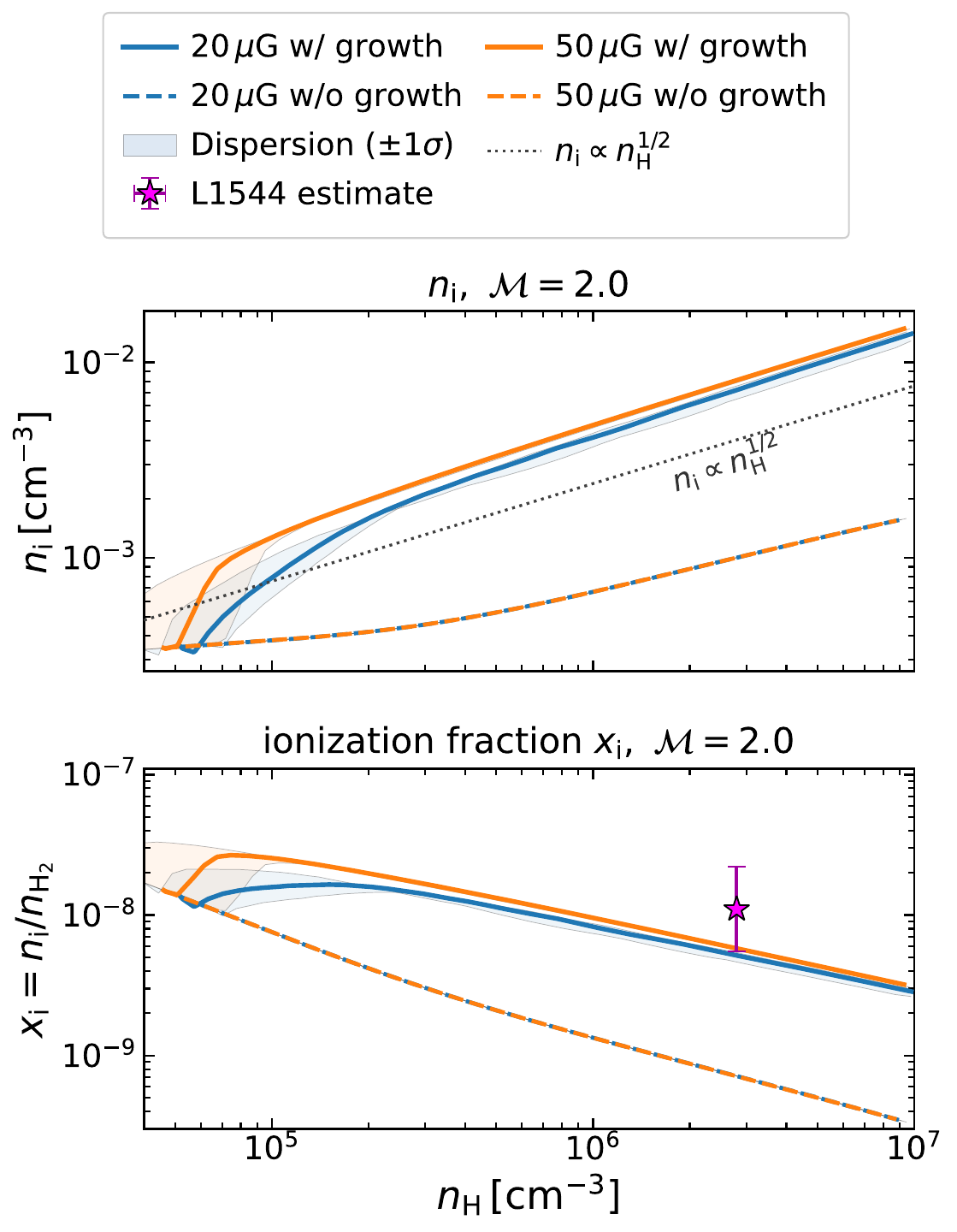}
  \caption{Dependence of the ion number density $n_{\mathrm{i}}$ (top) and the ionization fraction $x_{\mathrm{i}}\equiv n_{\mathrm{i}}/n_{\mathrm{H}_2}$ (bottom) on the number density of hydrogen nuclei $n_{\mathrm{H}}$. The blue and orange lines represent the weak- and strong-magnetic-field models, respectively. In the legend, solid and dashed lines are labeled ``w/ growth'' and ``w/o growth,'' respectively. The shaded bands indicate the dispersion ranges ($\pm1\sigma$), and the gray dotted line in the top panel shows the reference scaling $n_{\mathrm{i}}\propto n_{\mathrm{H}}^{1/2}$. The magenta star marks the reference estimate for L1544, $x_{\mathrm{i}}\sim1.1\times10^{-8}$ at $n_{\mathrm{H}_2}=1.4\times10^{6}\,\mathrm{cm^{-3}}$, with the molecular-hydrogen density converted to $n_{\mathrm{H}}=2n_{\mathrm{H}_2}$ for the horizontal coordinate. The vertical error bar indicates a factor-of-2 range around the reference ionization fraction.}
  \label{fig-io-deg}
\end{figure}

Figure~\ref{fig-io-deg} shows the ion number density $n_{\mathrm{i}}$ and the ionization fraction $x_{\mathrm{i}}$ as functions of the number density of hydrogen nuclei $n_{\mathrm{H}}$.

Dust growth depletes small dust grains and thereby reduces the adsorption of charged particles onto dust surfaces, producing higher $n_{\mathrm{i}}$ and $x_{\mathrm{i}}$ than in the models without dust growth. At high densities, $n_{\mathrm{i}}$ approaches the scaling $n_{\mathrm{i}}\propto n_{\mathrm{H}}^{1/2}$, indicating that gas-phase ionization and recombination become increasingly important. Although $x_{\mathrm{i}}$ decreases with increasing density, dust growth raises it from $\sim10^{-9}$ to $\sim10^{-8}$ around $n_{\mathrm{H}}\sim10^6\,\mathrm{cm^{-3}}$.

For the representative central density of L1544, $n_{\mathrm{H}_2}\simeq1.4\times10^6\,\mathrm{cm^{-3}}$ \citep{2002ApJ...569..815T}, corresponding to $n_{\mathrm{H}}\simeq2.8\times10^6\,\mathrm{cm^{-3}}$, the standard density--ionization relation gives $x_{\mathrm{i}}\simeq1.1\times10^{-8}$ \citep{1989ApJ...345..782M}.

If this value is representative of L1544, our results suggest that dust growth is also required from the viewpoint of the ionization fraction. We note, however, that detailed chemical models predict a central electron fraction of order $10^{-9}$ \citep{2002ApJ...565..344C,2021A&A...656A.109R}, so some uncertainty remains in this comparison.

\subsection{Magnetic-field Dependence of the Infrared Scattering Albedo}
\label{sec:infrared_scattering}
\begin{figure*}[!t]
  \centering
  \includegraphics[width=0.94\textwidth]{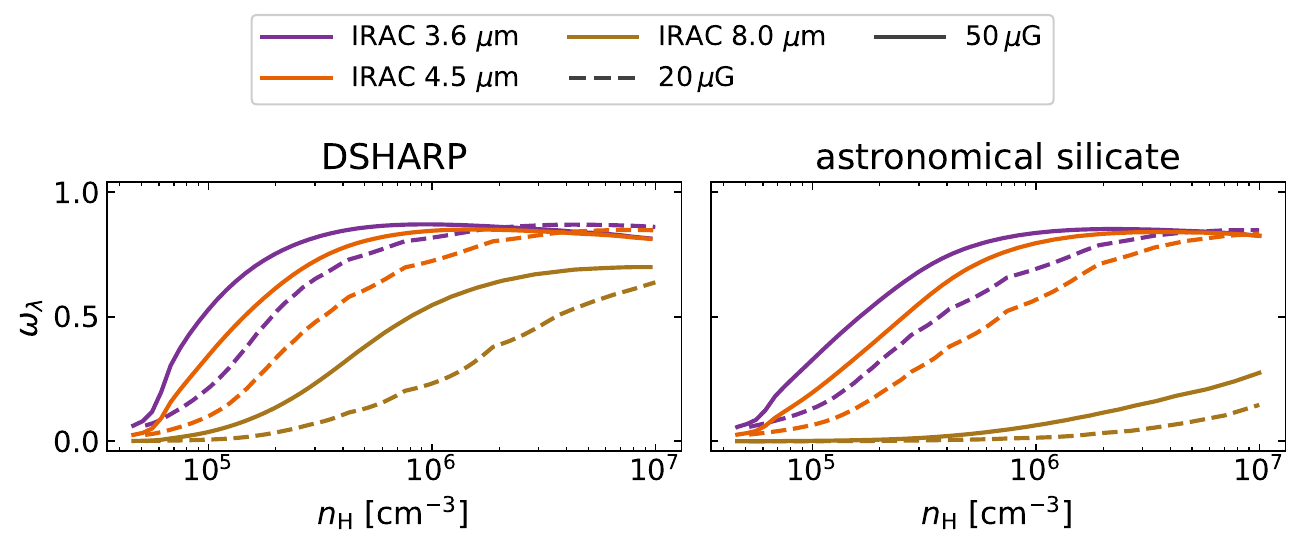}
  \caption{Infrared scattering properties computed from the evolved dust size distributions using OpTool \citep{2021ascl.soft04010D}. The single-scattering albedo $\omega_\lambda$ is shown as a function of the number density of hydrogen nuclei $n_{\mathrm{H}}$. The left and right panels show the results for the DSHARP mixture and astronomical silicate compositions, respectively. The purple, orange, and brown lines correspond to the IRAC $3.6$, $4.5$, and $8.0\,\mu\mathrm{m}$ bands, respectively. The dashed and solid lines represent the weak- and strong-magnetic-field models, respectively.}
  \label{fig:optool_albedo_ksca}
\end{figure*}

Figure~\ref{fig:optool_albedo_ksca} shows the infrared scattering properties computed from the dust size distributions obtained in the weak- and strong-magnetic-field models.

The single-scattering albedo is defined as
\begin{equation}
\omega_\lambda =
\frac{\kappa_{\mathrm{sca}}(\lambda)}
{\kappa_{\mathrm{abs}}(\lambda)+\kappa_{\mathrm{sca}}(\lambda)} ,
\end{equation}
where $\kappa_{\mathrm{abs}}$ and $\kappa_{\mathrm{sca}}$ are the absorption and scattering opacities, respectively.

For both dust compositions, $\omega_\lambda$ rises at lower densities in the strong-magnetic-field model than in the weak-magnetic-field model because the slower contraction allows dust growth to proceed further by the time the gas reaches a given density. This difference is particularly clear at $3.6$ and $4.5\,\mu\mathrm{m}$, indicating that a stronger magnetic field shifts the onset of efficient ``coreshine''-band scattering toward lower densities. At sufficiently high densities, however, both magnetic-field models attain high albedos of order $\omega_\lambda\sim0.8$ in these bands.

The DSHARP mixture and astronomical silicate show broadly similar trends at $3.6$ and $4.5\,\mu\mathrm{m}$, whereas their behavior differs substantially at $8.0\,\mu\mathrm{m}$. The $8.0\,\mu\mathrm{m}$ albedo rises more slowly because efficient scattering at this wavelength requires larger dust grains. At high densities, it reaches $\omega_\lambda\sim0.7$ for the DSHARP mixture but remains much lower for astronomical silicate because absorption near the $10\,\mu\mathrm{m}$ silicate feature suppresses the scattering albedo.

\section{Discussion}
\label{sec:4}
Most previous dust growth calculations have adopted fixed physical conditions or idealized collapse models, whereas dense cores commonly form within filamentary molecular clouds \citep{Andre2014,Pineda2023}. Here, we followed time-dependent density and magnetic-field histories extracted from three-dimensional non-ideal MHD simulations of core formation through filament fragmentation. The central result is that the magnetic field regulates dust evolution through the collapse timescale: the stronger magnetic field delays contraction, giving dust grains more time to grow at a given density and thereby enhancing the depletion of very small dust grains. This establishes a direct link between magnetically regulated core formation and the dust grain microphysics that controls ionization and non-ideal MHD effects.

Previous collapse calculations have also shown that small-grain
depletion can increase the ambipolar resistivity.
\citet{2020A&A...643A..17G} and \citet{2023MNRAS.518.3326L}
showed that ambipolar-diffusion-induced grain drift, which depends
on grain size, can promote the removal of small grains; in the latter
study, this depletion is enhanced for stronger magnetic fields.
In these studies, however, the collapse history itself is not dynamically
regulated by the magnetic field.
In contrast, our stronger-field model contracts more slowly, providing
more time for coagulation at a given density and thereby enhancing
dust growth and small-grain depletion.
Thus, while the resulting increase in ambipolar resistivity is
qualitatively consistent with these previous studies, the physical
pathway by which the magnetic field affects dust evolution is different
in the present work.

The depletion of very small dust grains reduces both the adsorption of
charged particles onto dust surfaces and the contribution of charged
dust grains to the conductivity, leading simultaneously to higher
$x_{\mathrm{i}}$, larger $\eta_{\mathrm{A}}$, and faster ion--neutral drift.
This sensitivity to the small-grain population is consistent with
previous studies showing that very small dust grains are important
regulators of the charge distribution and non-ideal MHD resistivities
\citep{2021MNRAS.505.5142Z,2022ApJ...934...88T}. If the estimate $x_{\mathrm{i}}\simeq1.1\times10^{-8}$ for L1544 is representative \citep{1989ApJ...345..782M,2002ApJ...569..815T}, our results suggest that dust growth is required to reproduce its ionization fraction; we note, however, that detailed chemical models predict a central electron fraction of order $10^{-9}$ \citep{2002ApJ...565..344C,2021A&A...656A.109R}. Dust growth also raises $v_{\mathrm{drift}}$ to several $\mathrm{m\,s^{-1}}$ in the weak-magnetic-field model and approximately $10\,\mathrm{m\,s^{-1}}$ in the strong-magnetic-field model, but these values remain below the observationally suggested range of $30$--$100\,\mathrm{m\,s^{-1}}$ \citep{2026arXiv260522541A}. Reproducing the inferred drift velocity may therefore require stronger magnetic fields than those considered here.

The magnetically regulated dust growth history also has observable consequences for infrared scattering. The micron-sized dust grains produced by dense-core conditions are comparable to those invoked in radiative-transfer models of ``coreshine'' \citep{2013A&A...559A..60A} and are consistent with its interpretation as scattering by grown dust grains \citep{2010A&A...511A...9S,2010Sci...329.1622P}. Because dust growth proceeds further in the strong-magnetic-field model at a given density, the $3.6$ and $4.5\,\mu\mathrm{m}$ albedos rise at lower densities than in the weak-magnetic-field model, although both models eventually attain high albedos in these bands. The composition dependence is most pronounced at $8.0\,\mu\mathrm{m}$, where absorption associated with the silicate feature suppresses the albedo of astronomical silicate relative to the DSHARP mixture. A high albedo alone does not determine the observed surface brightness, which also depends on the radiation field, density structure, optical depth, and viewing geometry; direct comparison with ``coreshine'' observations therefore requires radiative-transfer calculations using the evolved dust size distributions.

The main theoretical limitation is that the present model remains a post-processed one-zone calculation: changes in the ionization balance and ambipolar resistivity caused by dust growth do not feed back on the gas dynamics or magnetic-field evolution. 
In addition, all collisions between dust grains are assumed to result in coagulation, whereas fragmentation, bouncing, and electrostatic repulsion may suppress dust growth; Coulomb barriers can be particularly important for like-charged small dust grains \citep{2009ApJ...698.1122O}.

To assess the possible impact of fragmentation, we compare the calculated
collision velocities with representative fragmentation thresholds.
Even in the model showing the strongest dust growth, the equal-size collision
velocities remain below the $\sim300\,{\rm m\,s^{-1}}$ threshold for
ice-coated aggregates composed of $0.1\,\mu{\rm m}$ monomers
\citep{2009A&A...502..845O,2023MNRAS.518.3326L}.
In contrast, the much lower $\sim15\,{\rm m\,s^{-1}}$ threshold for
bare-silicate aggregates can be exceeded in our calculations, suggesting that
fragmentation may become important for bare-silicate grains.

Charged dust grains may also couple to magnetic-field lines and participate in ambipolar drift \citep{1996ApJ...468..749C,2020A&A...643A..17G,2020A&A...641A..39S}. A fully coupled multidimensional calculation including these processes is therefore needed to quantify their feedback on core formation.

Observational comparisons introduce additional uncertainties. Estimates of ion--neutral drift depend on tracer selection, line-of-sight averaging, projection, and radiative-transfer effects \citep{2026ApJ...999...79F}, while infrared surface brightness requires consistent treatment of illumination and geometry. Forward modeling of molecular-line emission and dust scattering from coupled non-ideal MHD and dust evolution calculations will be necessary for direct comparison with observations. Nevertheless, the present results robustly show that the collapse history set by the magnetic field controls dust growth and thereby links dust size evolution, ionization, ambipolar diffusion, and infrared scattering in dense cores.

\FloatBarrier
\restartappendixnumbering
\makeatletter
\@two@colfalse
\@two@col@appfalse
\appendix
\@two@coltrue
\@two@col@apptrue
\makeatother
\raggedbottom
\section{Dust Relative Velocity}

Figure~\ref{fig:appendix_relvel_temp} shows the relative velocity between same-size dust grains as a function of grain size for the initial condition of the strong-magnetic-field model.
For small grains, the relative velocity is dominated by Brownian motion and decreases with increasing grain size.
In contrast, the turbulent relative velocity increases with grain size because larger grains are less tightly coupled to the gas and cannot fully follow turbulent fluctuations.
The total relative velocity therefore transitions from Brownian-dominated at small grain sizes to turbulence-dominated at large grain sizes.

\begin{inlinecolumnfigure}
    \centering    \includegraphics[scale=0.5]{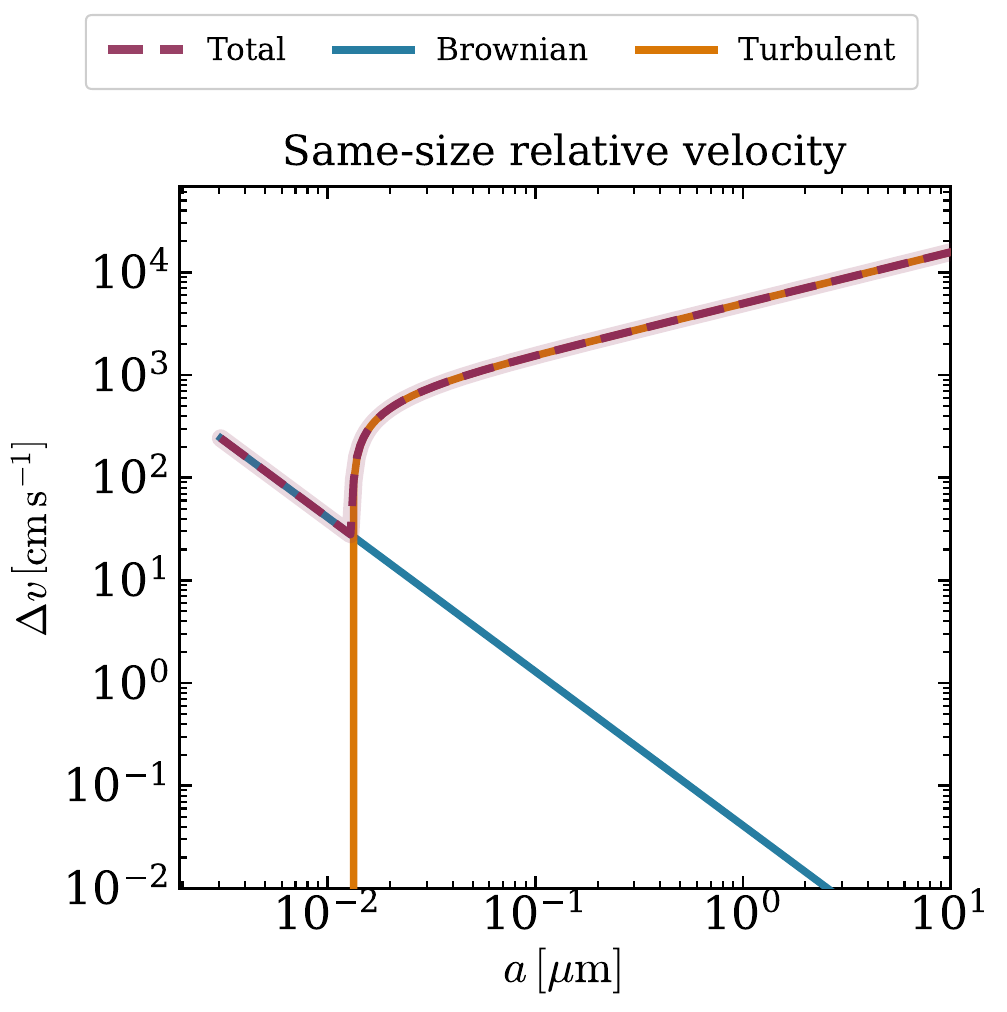}
    \caption{
    Relative velocity between same-size dust grains as a function of
    dust size at the initial condition of the strong-magnetic-field model.
    The blue and orange lines show the contributions from Brownian motion
    and turbulence, respectively. The magenta dashed line shows the total
    relative velocity.
    }
    \label{fig:appendix_relvel_temp}
\end{inlinecolumnfigure}

\section{Definitions of Weighted Mean Dust Sizes}
\label{def:mean_size}

To characterize the evolving dust size distribution, we use four weighted mean dust sizes: the number-, radius-, area-, and mass-weighted means. 
Here, $a$ denotes the dust size, $t$ denotes the time, and $n(a,t)\,da$ denotes the number density of dust grains with sizes between $a$ and $a+da$ at time $t$. 
The integrals below are evaluated over the dust size range used in the dust evolution calculation. 
With these definitions, the number-weighted mean size is given by
\begin{equation}
\langle a \rangle_{\mathrm{num}}
=
\frac{\int a\,n(a,t)\,da}
{\int n(a,t)\,da}.
\end{equation}

The radius-weighted mean size is defined as
\begin{equation}
\langle a \rangle_{\mathrm{rad}}
=
\frac{\int a\,[a\,n(a,t)]\,da}
{\int a\,n(a,t)\,da}
=
\frac{\int a^{2} n(a,t)\,da}
{\int a\,n(a,t)\,da}.
\end{equation}

The area-weighted mean size is defined as
\begin{equation}
\langle a \rangle_{\mathrm{area}}
=\frac{\int a^{3} n(a,t)\,da}
{\int a^{2} n(a,t)\,da},
\end{equation}
where $\pi a^2$ is the geometrical cross-sectional area of a dust grain.

The mass-weighted mean size is defined as
\begin{equation}
\langle a \rangle_{\mathrm{mass}}
=
\frac{\int a\,m(a)\,n(a,t)\,da}
{\int m(a)\,n(a,t)\,da},
\end{equation}
\par\vspace*{0.5\baselineskip}
\noindent where $m(a)$ is the mass of a dust grain with size $a$. 
Assuming a fixed material density $\rho_{\mathrm{mat}}$, the dust mass is
\begin{equation}
m(a)=\frac{4}{3}\pi \rho_{\mathrm{mat}} a^{3},
\end{equation}
where $\rho_{\mathrm{mat}}$ is the material density of the dust grains. 
The mass-weighted mean size can therefore be written as
\begin{equation}
\langle a \rangle_{\mathrm{mass}}
=
\frac{\int a^{4} n(a,t)\,da}
{\int a^{3} n(a,t)\,da}.
\end{equation}

In the numerical implementation, these quantities are evaluated as discrete sums over the dust size bins.

\begin{acknowledgments}
We thank the anonymous referee for the insightful comments.
Numerical computations were carried out on the Cray XD2000 at the Center for Computational Astrophysics, National Astronomical Observatory of Japan.
This work was supported by JSPS KAKENHI Grant Number JP23K19073.
\end{acknowledgments}

\bibliographystyle{aasjournalv7}
\bibliography{export-bibtex}

\end{document}